# Towards A Resolution Of The Vacuum Energy Density Crisis

Robert L. Oldershaw
12 Emily Lane
Amherst, MA 01002
USA
rloldershaw@amherst.edu
011-413-549-1220

**Abstract:** The theoretical vacuum energy density estimated on the basis of the standard model of particle physics and very general quantum assumptions is 59 to 123 *orders of magnitude* larger than the measured vacuum energy density for the observable universe which is determined on the basis of the standard model of cosmology and empirical data. This enormous disparity between the expectations of two of our most widely accepted theoretical frameworks demands a credible and self-consistent explanation, and yet even after decades of sporadic effort a generally accepted resolution of this crisis has not surfaced. Very recently, however, a discrete self-similar cosmological paradigm based on the fundamental principle of discrete scale invariance has been found to offer a rationale for reducing the vacuum energy density disparity by at least 115 orders of magnitude, and possibly to eliminate the vacuum energy density crisis entirely.

## 1. Introduction

The vacuum energy density crisis is perhaps the most dramatic manifestation of the fact that physics is still very much a "divided house", with quantum physics ruling the microcosm and general relativity dominating the macrocosm. Within their own domains, quantum physics and general relativity are thought to be on very strong empirical and theoretical footing, so it is very disconcerting to find that these two foundational frameworks strongly contradict each other when they meet at the "intersection" of the vacuum energy density.

Nobelist Frank Wilczek (2001) has characterized the situation as follows. "We do not understand the disparity. In my opinion, it is the biggest and most profound gap in our current understanding of the physical world. … [The solution to the problem] might require inventing entirely new ideas, and abandoning old ones we thought to be well-established. … Since vacuum energy density is central to both fundamental physics and cosmology, and yet poorly understood, experimental research into its nature must be regarded as a top priority for physical science."

The vacuum energy density (VED) is generally viewed as a fundamental property of the cosmos whose magnitude should not depend upon whether we choose subatomic, astronomical or cosmological methods to assess its value. As Wilczek notes, the fact that we get such wildly differing values when using subatomic and cosmological analyses means that there must be a serious flaw in the reasoning involved in one or both of those analyses. This is an enigma that pertains to the entire discipline of physics. Since astronomical and astrophysical observations and analyses play a major role in arriving at the most consistent empirically-based determinations of the vacuum energy density, the VED crisis is highly relevant to astrophysics.

## 2. Technical Issues

### 2.1 The Vacuum Energy Density of High Energy Physics

Excellent reviews of the basic physics and the more technical matters involved in determining the vacuum energy density in the contexts of high energy physics and cosmology can be found in papers by Carroll, Press and Turner (1992) and Weinberg (1989). According to general assumptions of quantum physics and quantum field theory, the high energy physics (HEP) vacuum contains many fields and can be modeled in terms of quantum harmonic oscillations occurring at each point of the fields constituting the vacuum. To evaluate the HEP vacuum energy density one assumes that there is a particle in each unit volume of the vacuum, which can be defined as the cube of the relevant Compton wavelength. A cutoff to the appropriate energies and wavevectors of the vacuum fluctuations is required in order to avoid an "ultraviolet divergence" and an infinite vacuum energy density. The most common assumption regarding this cutoff is that the conventional Planck scale defines the most appropriate ultraviolet cutoff scale for calculating the HEP vacuum energy density. In this case,

$$\rho_{hep} = \mathfrak{M}^4 c^3/h^3 = 2.44 \text{ x } 10^{91} \text{ g/cm}^3 , \qquad (1)$$

where $\rho_{hep}$ is the vacuum energy density of high energy physics, $\mathfrak{M}$ is the Planck mass, c is the velocity of light and h is Planck's constant.

Different $\rho_{hep}$ values have been estimated depending on the particular theoretical method employed, the identification of the relevant contributions of different particles/fields, the choice of an appropriate ultraviolet cutoff, and the possibility of various cancellation mechanisms.

Wilczek (2001) notes that one can estimate a $\rho_{hep}$ of about $10^{108}$ ev$^4$ based on the standard quantum gravity/Planck scale cutoff method, a $\rho_{hep}$ of about $10^{96}$ ev$^4$ based on unified gauge symmetry breaking, and a $\rho_{hep}$ as small as $10^{44}$ ev$^4$ if "low-energy supersymmetry enforces big cancelations." Various authors have noted that $\rho_{hep}$ values can range from about $10^{92}$ g/cm$^3$ down to about $10^{30}$ g/cm$^3$. In light of this uncertainty, one might be permitted to ask the following impudent questions. If the standard model is the "towering edifice" that many particle physicists claim it is, then why does it yield $\rho_{hep}$ estimates scattered over a range that is 60 orders of magnitude wide, and why do none of these estimates appear to be compatible with observational limits? For the remainder of this paper we will adopt the standard quantum gravity/Planck scale cutoff calculation of $\rho_{hep}$ because it employs the most common set of assumptions and can be viewed as the default method of determining the vacuum energy density of high energy physics.

### 2.2 The Cosmological Vacuum Energy Density

In the cosmological context, the calculation of $\rho_{cos}$ is a bit more straightforward (Padmanabhan, 2003). The critical density defining the dividing line between open and closed solutions of the standard cosmological model is:

$$\rho_{cr} = 3H^2/8\pi G = 1.88(h^2) \times 10^{-29} \text{ g/cm}^3 , \qquad (2)$$

where $\rho_{cr}$ is the critical density, H is the current value of the Hubble constant, G is the conventional Newtonian gravitational constant and $h \equiv H/100$ km sec$^{-1}$ Mpc$^{-1}$. Given that the cosmological constant ($\lambda$) is currently inferred to be $\approx 0.75\ \rho_{cr}$ , and constitutes a good approximation to the cosmological vacuum energy density, cosmologists find that:

$$\rho_{cos} \approx 0.75\ \rho_{cr} \sim 10^{-29}\ \text{g/cm}^3. \qquad (3)$$

**2.3 The Crisis**

Because a basic tenet of general relativity is that all forms of energy, including $\rho_{hep}$ , will contribute to the value of $\rho_{cos}$, we are confronted with the apparent empirical fact that something is seriously wrong with the enormous theoretical $\rho_{hep}$ value. It seems virtually inconceivable that $\rho_{cos}$ could be *many* orders of magnitude larger than the observed value. On the other hand, a very high $\rho_{hep}$ seems almost mandatory from the general point of view of quantum physics. Moreover, a large $\rho_{hep}$ seems to be required in at least two additional critical areas of physics. Firstly, a high $\rho_{hep}$ would seem to be a necessary prediction of the Higgs mechanism, and associated Higgs field, which is hypothesized to give subatomic particles their mass values and is a cornerstone of the standard model of particle physics. Secondly, the inflationary scenario which is crucial to the standard model of cosmology also requires a very large value of the vacuum energy density in the early universe.

In sum, Wilczek (2001) is fully justified in saying that the disparity between $\rho_{hep}$ and $\rho_{cos}$ indicates that there must be one or more fundamental errors in the standard models of particle physics and cosmology. His admonition that new ideas must be considered and that assumptions that were previously regarded as virtually sacrosanct may need to be revised or discarded can be seen as sufficient motivation for considering the discrete self-similar paradigm's radical approach to resolving the vacuum energy density crisis.

## 3. A New Approach To The Crisis

### 3.1 The Discrete Self-Similar Paradigm

The arguments presented below are based on the self-similar cosmological paradigm (SSCP) (Oldershaw, 1987; 1989a, b; 2002; 2007) which has been developed over a period of more than 30 years, and can be unambiguously tested via its definitive predictions (Oldershaw, 1987; 2002) concerning the nature of the galactic dark matter. Briefly, the discrete self-similar paradigm focuses on nature's fundamental organizational principles and symmetries, emphasizing nature's *intrinsic hierarchical organization* of systems from the smallest observable subatomic particles to the largest observable superclusters of galaxies. The new discrete fractal paradigm also highlights the fact that nature's global hierarchy is *highly stratified*. While the observable portion of the entire hierarchy encompasses nearly 80 orders of magnitude in mass, three narrow mass ranges, each extending for only about 5 orders of magnitude, account for ≥ 99% of all mass observed in the cosmos. These dominant mass ranges: roughly $10^{-27}$ g to $10^{-22}$ g, $10^{28}$ g to $10^{33}$ g and $10^{38}$ g to $10^{43}$ g, are referred to as the *Atomic, Stellar and Galactic Scales*, respectively. The cosmological Scales constitute the discrete self-similar scaffolding of the observable portion of nature's quasi-continuous hierarchy. At present the number of Scales cannot be known, but for reasons of natural philosophy it is tentatively proposed that there are a denumerably infinite number of cosmological Scales, ordered in terms of their intrinsic ranges of space, time and mass scales. A third general principle of the new paradigm is that the *cosmological Scales are rigorously self-similar* to one another, such that for each class of fundamental particles, composite systems or physical phenomena on a given Scale there is a corresponding class of particles, systems or phenomena on all other cosmological Scales. Specific self-similar analogues from different Scales have rigorously analogous morphologies, kinematics and dynamics. When the general self-similarity among the discrete Scales is *exact*, the paradigm is referred to as discrete scale relativity (DSR) (Oldershaw, 2007) and nature's global space-time geometry manifests a new universal symmetry principle: *discrete scale invariance*.

Based upon decades of studying the scaling relationships among analogue systems from the Atomic, Stellar and Galactic Scales (Oldershaw, 1987; 1989a, b; 2001; 2002; 2007), a close approximation to nature's self-similar Scale transformation equations for the length (L), time (T) and mass (M) parameters of analogue systems on neighboring cosmological Scales Ψ and Ψ-1, *as well as for all dimensional constants*, are as follows.

$$L_\Psi = \Lambda L_{\Psi-1} \quad (4)$$

$$T_\Psi = \Lambda T_{\Psi-1} \quad (5)$$

$$M_\Psi = \Lambda^D M_{\Psi-1} \quad (6)$$

The self-similar scaling constants Λ and D have been determined empirically and are equal to ≅ $5.2 \times 10^{17}$ and ≅ 3.174, respectively (Oldershaw, 1989a, b). The value of $\Lambda^D$ is $1.70 \times 10^{56}$. Different cosmological Scales are designated by the discrete index Ψ (≡ …, -2, -1, 0, 1, 2, …) and the Atomic, Stellar and Galactic Scales are usually assigned Ψ = -1, Ψ = 0 and Ψ = +1, respectively.

The fundamental self-similarity of the SSCP and the recursive character of the discrete scaling equations suggest that nature is an infinite discrete fractal, in terms of its morphology, kinematics and dynamics. The underlying principle of the paradigm is discrete scale invariance and the physical embodiment of that principle is the discrete self-similarity of nature's physical systems. Perhaps the single most thorough and accessible resource for exploring the SSCP is the author's website (Oldershaw, 2001).

### 3.2 A Revised Scaling For Gravitation

Because the discrete self-similar scaling of the new paradigm applies to *all dimensional parameters*, the Scale transformation equations also apply to dimensional "constants." It has been shown (Oldershaw, 2007) that the gravitational coupling constant $G_\Psi$ scales as follows.

$$G_\Psi = (\Lambda^{1-D})^\Psi G_0 \quad , \qquad (7)$$

where $G_0$ is the conventional Newtonian gravitational constant. Eq. (7) results from the $L^3/MT^2$ dimensionality of $G_\Psi$ and the self-similar scaling rules embodied in Eqs. (4) - (6). Therefore the Atomic Scale value $G_{-1}$ is $\Lambda^{2.174}$ times $G_0$ and equals $\cong 2.18 \times 10^{31} cm^3/g\ sec^2$.

The value of the gravitational coupling constant has been tested on a variety of size scales, but it has never been empirically measured *within* an Atomic Scale system. To be perfectly clear on this point, the distinction between the appropriateness of using $G_0$ or $G_{-1}$ as the correct gravitational coupling constant is less determined by *size* scales than by whether the region of interest is *within* an Atomic Scale system, or *exterior* to Atomic Scale systems. The possibility that the Atomic Scale gravitational coupling factor is on the order of $10^{38}$ times larger than its counterpart within a Stellar Scale system has recently found support in successful retrodictions of the proton mass and radius using the geometrodynamic form of Kerr-Newman solutions to the Einstein-Maxwell equations (Oldershaw, 2010a), and in the discovery of a natural and compelling explanation for the meaning of the fine structure constant (Oldershaw, 2010b).

### 3.3 A Revised HEP Vacuum Energy Density

The conventional Planck scale is based on the use of $G_0$ to determine the numerical values of the Planck mass, length and time. However, if the revised scaling for gravitation proposed by the SSCP is correct, then a revised Planck scale based on $G_{-1}$ is necessary and the revision yields the following values.

$$\text{Planck length} = (\hbar G_{-1}/c^3)^{1/2} = 2.93 \times 10^{-14}\ \text{cm} \approx 0.4\ \text{proton radius} \qquad (8)$$

$$\text{Planck mass} = (\hbar c/G_{-1})^{1/2} = 1.20 \times 10^{-24}\ \text{g} \approx 0.7\ \text{proton mass} \qquad (9)$$

$$\text{Planck time} = (\hbar G_{-1}/c^5)^{1/2} = 9.81 \times 10^{-25}\ \text{sec} \approx 0.4\ (\text{proton radius}/c) \qquad (10)$$

When the revised Planck mass ($\mathfrak{M}_{-1}$) of $1.20 \times 10^{-24}$ g is substituted for the conventional Planck mass in Eq. (1), then

$$\rho_{hep} = (\mathfrak{M}_{-1})^4\, c^3 / h^3 = 2.3 \times 10^{14}\ \text{g/cm}^3 \qquad (11)$$

Within the context of the SSCP, the value of $\rho_{hep}$ has been *revised downward by about 77 orders of magnitude*, roughly from $\sim 10^{91}$ g/cm$^3$ to $\sim 10^{14}$ g/cm$^3$. This huge decrease in $\rho_{hep}$ is a direct result of the SSCP's contention that the coupling between matter and space-time geometry within Atomic Scale systems is $\approx 3 \times 10^{38}$ times stronger than is *conventionally assumed*.

### 3.4 A Revised Cosmological Vacuum Energy Density

When cosmologists evaluate Eq. (2) in the conventional manner, they use $G_0$ because this is assumed to be the correct gravitational coupling factor in *any* context. However, according to the discrete gravitational scaling of the SSCP the Galactic Scale value $G_1$, which $\approx \Lambda^{-2.174} G_0$ or $\approx (3.06 \times 10^{-39})(G_0)$, is required for a more accurate evaluation of Eq. (2) at a Scale that is clearly "higher" than the Stellar Scale. Using $G_1$ as the appropriate gravitational coupling factor, we have

$$\rho_{cos} = 3H^2/8\pi G_1 = 6.14\,(h^2) \times 10^9 \text{ g/cm}^3 \qquad (12)$$

## 4. Towards A Resolution Of The Crisis

Within the context of the SSCP, we have found a $\rho_{cos}$ that is $\sim 10^{38}$ times larger than the conventional $\rho_{cos}$. When we combine the SSCP's reduction of $\sim 10^{77}$ in $\rho_{hep}$ with the SSCP's increase of $\sim 10^{38}$ in $\rho_{cos}$, the original 120 orders of magnitude disparity between $\rho_{hep}$ and $\rho_{cos}$ is reduced by about 115 orders of magnitude to a residual disparity of approximately 4.57 orders of magnitude. Because the theoretical estimate of $\rho_{hep}$ can vary by upwards of 60 orders of magnitude, technically there is no longer any guarantee that a *bona fide* vacuum energy density disparity still exists, if the SSCP's proposed scaling for gravitation is valid.

Even if one accepts the new self-similar cosmological paradigm and its discrete gravitational scaling, there are two issues that must be settled before we know whether the vacuum energy density crisis has been reduced from a disparity of ~ $10^{120}$ to a residual disparity of ~ $10^5$, or whether the crisis has been entirely removed by the SSCP.

(1) A more definitive theoretical prediction of $\rho_{hep}$ is required. Ideally that prediction should be accurate to within a factor of ± 3.

(2) There are reasonable arguments for using $G_1$ in Eq (2), but some might propose that using $G_2$ is a more appropriate choice of gravitational coupling factors in this context. If $G_2$ were the correct choice, then we would have the highly unusual problem of having $\rho_{cos} >> \rho_{hep}$! Such a result would be difficult to understand and so it is assumed here that the arguments favoring $G_1$ are more compelling than those for $G_2$. Resolving this issue requires a more thorough theoretical analysis of the subtleties involved in evaluating $\rho_{cos}$ within the context of the discrete self-similar paradigm, and will be discussed in a forthcoming paper.

At present it can only be claimed that the SSCP offers the *potential* for resolving the vacuum energy density crisis. However, given the extraordinary magnitude, seriousness and persistence of the VED disparity, the SSCP's potential solution can be viewed as a source of relief and encouragement. If the SSCP does represent a significant advance in efforts to unify the "divided house" of physics, then the fundamental assumptions of quantum physics and relativistic physics need to be reassessed *ab initio*.

## REFERENCES

Carroll, S. M., Press, W. H. & Turner, E. L. 1992, Ann. Rev. Astron. Astrophys. 30, 499

Oldershaw, R. L. 1987, Astrophys. J. 322, 34

Oldershaw, R. L. 1989a, Internat. J. Theor. Phys. 28, 669

Oldershaw, R. L. 1989b, Internat. J. Theor. Phys. 28, 1503

Oldershaw, R. L. 2001, http://www.amherst.edu/~rloldershaw

Oldershaw, R. L. 2002, Fractals 10, 27

Oldershaw, R. L. 2007, Astrophys. Space Sci. 311, 431. DOI: 10.107/s10509-007-9557-x (arXiv:physics/0701132v3)

Oldershaw, R. L. 2010a, J. of Cosmology 6, 1341 (arXiv:astro-ph/0701006v2)

Oldershaw, R. L. 2010b, *Chaos, Solitons & Fractals* in review, (arXiv:physics/07083501)

Padmanabhan, T. 2003, Phys. Rep. 380, 235

Weinberg, S. 1989, Rev. Mod. Phys. 61, 1

Wilczek, F. 2001, Int.J.Mod.Phys. A16, 1653 (arXiv:hep-ph/0101187)